\documentstyle[11pt,fullpage,doublespace,epsf]{article}
 \DeclareMathSizes{11}{19}{13}{9}   

\makeatletter
\@addtoreset{equation}{section}
\makeatother

\begin{document}
   
  \title{Computational approach to finite size and shape effects in iron nanomagnets}
\author{Michael McGuigan\\James Davenport\\Computational Science Center\\Brookhaven National Laboratory\\Upton NY 11973\\James Glimm\\
Stony Brook University\\mcguigan@bnl.gov\\jdaven@bnl.gov\\glimm@ams.sunysb.edu}
\date{}
\maketitle

\begin{abstract}We develop and validate a computational approach to nanomagnets. It is built on the spin wave approximation to a Heisenberg ferromagnet whose parameters   can be calculated from a first principles theory (e.g. density functional theory). The method can be used for high throughput analysis of a variety of nanomagnetic materials. We compute the dependence of the magnetization of an iron nanomagnet on temperature, size and shape. The approach is applied to nanomagnets in the range of  432 atoms to 59 million atoms, a size which is several orders of magnitude beyond the scalability of density functional theory.
\end{abstract}

\section{Introduction}

Nanomagnets are promising materials for high density data storage devices, as they represent a bit by a single dot or magnetic domain. In such systems data storage densities greater than $10^{12}\rm{\,bits }/\rm{in}^2$ may be possible while being thermally stable at room temperature \cite{Haast}.
Important design choices for such systems are the material composition, the number of atoms, and the shape of the dot. All of these choices can strongly affect the magnetic properties of the system. However, an experimental search of different combinations is time consuming. Thus, it is important to investigate efficient computational approaches to calculating  magnetic properties. One might expect such computational approaches to reduce the cost of bringing nanomagnetic storage devices to market.

A significant effort is devoted to the calculation of magnetic properties of materials from first principles. These studies typically describe zero temperature properties of a small number of atoms (less than a thousand). A effective spin model with parameters taken from such a first principle model can be simulated to  predict the behavior of larger systems at higher temperatures. 

Thus from both fundamental and practical points of view, an  efficient computational approach to magnets is important. One such approach is given by the Heisenberg model with the Hamiltonian:
\begin{equation}
H =  - \sum\limits_{i,j}J(i-j) {\vec S_i  \cdot \vec S_j }  + g\mu_B\vec B_{\rm{ext}}  \cdot \sum\limits_i {\vec S_i } + H_{\rm{crystal-aniso}}, 
\end{equation}
where $\vec S_i$ is a three component spin at lattice site  $i = (i_1 ,i_2 ,i_3 )$ satisfying the O(3) invariant constraint
\[
S_i^{x2}  + S_i^{y2}  + S_i^{z2}  = S(S+1).
\]
In the above, $g$ is the Lande g-factor, $\mu_B$ the Bohr magneton, $\vec B_{\rm{ext}}$ an external magnetic field and $S$ is the spin of the Heisenberg model. The first term is called the exchange Hamiltonian, the second the Zeeman Hamiltonian, and the third the crystal anisotropy. In this paper we shall restrict ourselves to the first term only. It provides a good description of the magnetization curve in zero magnetic field. We shall return to the physical effects of the remaining terms in a future study. The number of lattice sites or spins will be denoted by $N = 2N_xN_yN_z$ where $N_x, N_y,N_z$ are the number of atoms along a bcc lattice dimension for a rectangular parallelepiped. For the quantum Heisenberg model the spins do not commute and obey an algebra for which $[S^x,S^y] = i  S^z$. 

The parameters $J$ are exchange constants which can be calculated from a first principles model. Although the Heisenberg model can be formulated for any material where the $J$ values are known, in this paper we study the quantum Heisenberg model with $S=1$ used to describe bcc iron, which is an important material from a practical \cite{Pierce} as well as a fundamental point of view. The experimental data for the ratio of the magnetization of bcc iron to its value at $T=0$ is found in \cite{Crangle} to four digit accuracy. We will extend our analysis to other materials in future work.

\begin{table}[h]
\begin{center}
\begin{tabular}{ |c|c|} \hline

$i-j$ & $J[mRy]$\\ \hline
$(\frac{1}{2}, \frac{1}{2},\frac{1}{2})$ & 1.97 \\
(1,0,0) & .623 \\
(1,1,0) & -.132 \\
$(\frac{3}{2}, \frac{1}{2}, \frac{1}{2})$ & -.166 \\
(1,1,1)     &  -.271 \\
(2,0,0)     &   .056 \\
$(\frac{3}{2}, \frac{3}{2}, \frac{1}{2})$  & -.028 \\
(2,1,0)  &   .051 \\
(2,1,1) &  -.033  \\
$(\frac{3}{2},\frac{3}{2}, \frac{3}{2})$ & .096 \\ \hline

\end{tabular} 
\end{center}
\caption{$J$ values from reference [4]}
\label{tab1}
\end{table}
For bcc iron the $J$ values have been calculated \cite{Ant1} and are listed in Table 1.
Here one $\rm{mRy}$ corresponds to $13.6056923$ meV and $157.887324$ Kelvin.
We plot these $J $ values as a function of $r=\left|{i-j}\right|$ in Fig. 1.
We can see that the couplings are ferromagnetic out to second nearest neighbor.

\begin{figure}[htbp]
 
   \centerline{\hbox{
   \epsfxsize=5.0in
   \epsffile{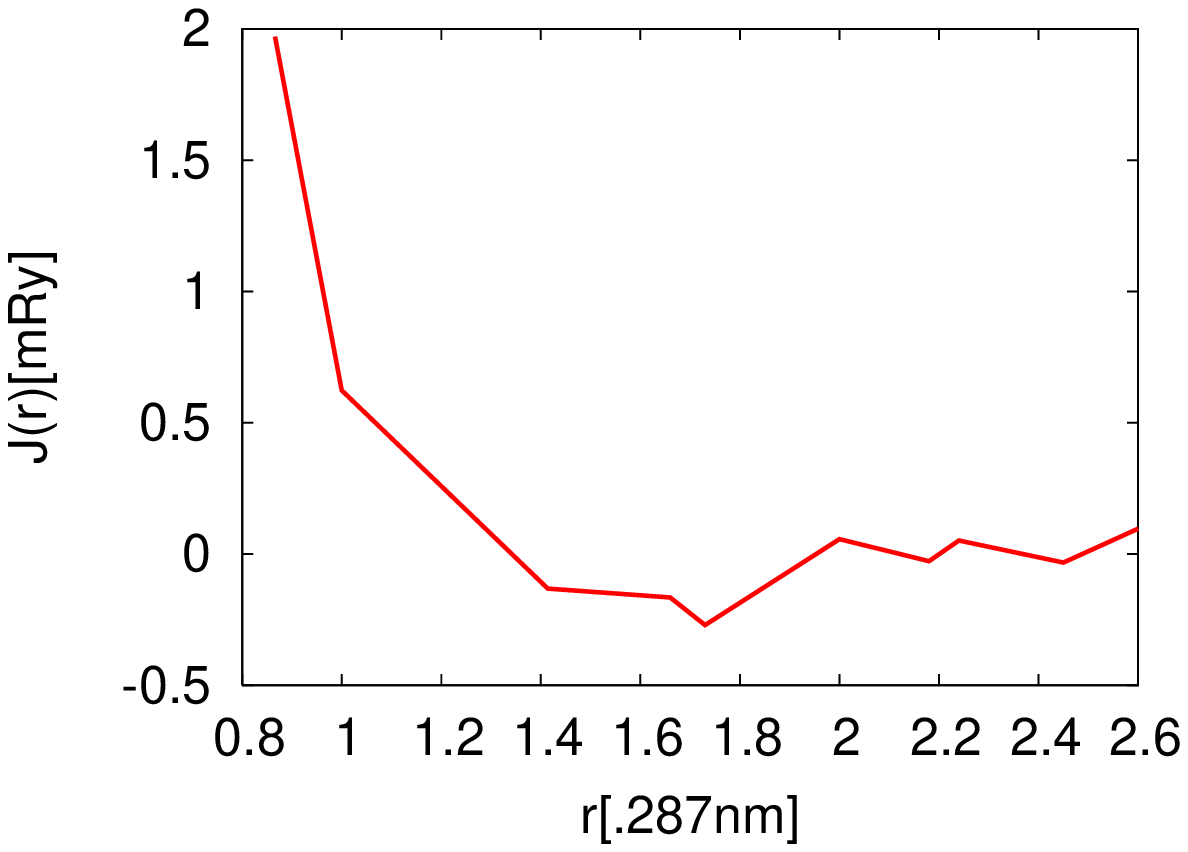}
     }
  }
  \caption{$J$ values from  [4] plotted as a function of $r=\left|{i-j}\right|$.}
           
  \label{fig1}
 
\end{figure}

Using somewhat different values for  $J$, the Curie temperature was estimated to be $1414 K$ in a mean field approximation and $950 K$ using the Random Phase approximation \cite{Pajda}. The latter is within 10 per cent of the observed value $1043K$. An even better estimate, obtained using  a local spin density approximation \cite{Ant2}  gives $1070 K$. 

The above results indicate that the Heisenberg model of ferromagnetism can be derived from first principles and gives a good description of ferromagnetism in bulk bcc iron. In the next two sections we apply this model to the description of bcc iron nanomagnets, which are of great experimental and theoretical interest.
 
This paper is organized as follows. In Section 1 we summarize our computational method. In Section 2 we compute the finite volume effects in iron nanomagnets. In Section 3 we compute the finite shape effects in iron nanomagnets at fixed volume. In Section 4 we present our main conclusions.

Our computational method can be summarized as follows:

(1) Determine the parameters $J$ of an effective Heisenberg model from a first principles zero temperature calculation and/or low temperature measurement.

(2) Choose a lattice geometry describing the nanomagnet. This will determine its size and shape. In this paper we will choose bcc iron and a parallelepiped lattice geometry, but other choices are possible.

(3) Apply the Holstein-Primakoff transformation to map the spin variables to a set of harmonic oscillators whose nearest neighbor interactions determine a lattice Laplacian.

(4) Solve for the eigenstates and eigenvalues of this lattice Laplacian.

(5) Use these eigenvalues to determine the magnon dispersion relation.

(6) The main computational step is to construct a sum over these eigenvalues using Bose-Einstein statistical mechanics to determine the magnetization, given by the expectation value $M = \langle \sum\limits_i S^z_i \rangle$ as a function of temperature. We define the thermal average by $\langle O\rangle= tr\left(Oe^{-H/kT})\right)$ for an operator $O$.

(7) Vary the temperature, size and shape of the nanomagnet to determine its range of magnetic properties.

(8) Compare the above derived magnetization curves against experiment.

We comment on the main steps (3) and (4). Since we consider only nearest neighbor terms, the Holstein-Primakoff transformation, which is a linearization of the full many body Hamiltonian, yields nearest neighbor coupling only, and a discrete version of the Laplace operator. The standard continuum Laplacian is an approximation which ignores atomic spacing and lattice geometry. The construction thus is applicable to arbitrary lattices, including lattices with defects and arbitrary shapes. For such problems, the eigenvalues of the Hamiltonian are solved numerically. For simplicity we consider here a rectangular geometry and an absence of defects. In this case the eigenvalues are known in closed form.

A similar approach can be found in the description of spin-waves in ribbon shaped iron nanoparticles in \cite{Crespo}. An earlier study of the finite size effects, based on the analysis of the spin-wave spectrum of Heisenberg spins was performed in \cite{Hend}.  The inclusion of higher order spin wave interactions can also be considered computationally \cite{Sun}. These become more important as one approaches the Curie temperature \cite{Kobler} and can included in full quantum monte carlo simulations of the quantum Heisenberg model \cite{Todo}.

\section{Finite size effects in bcc iron nanomagnets} 

The presence of the $T^{3/2}$ term in the empirical formulas for bcc iron indicate that quantized spin-waves or magnons and the associated Bloch law are physically very important \cite{Dyson1,Dyson2}.

The magnon or spin-wave appoach is most easily introduced using the transformation of spin variables of Holstein and Primakoff \cite{Holstein}. This transformation is given by
\[
\begin{array}{l}
 S_i^ +   = \sqrt {2S} f_i a_i  \\
 S_i^ -   = \sqrt {2S} a_i^ +  f_i  \\
 S_i^z  = S - a_i^ +  a_i  \\
 \end{array}
\]
where $a_i$ and $a_i^+$ satisfy the usual harmonic oscillator algebra and $f_i$, $S_i^+$, and $S_i^-$ are defined by:
\[
\begin{array}{l}
 f_i  = (1 - a_i^ +  a_i /2S)^{1/2}  \\
 S_i^ +   = S_i^x  + iS_i^y  \\
 S_i^ -   = S_i^x  - iS_i^y  \\
 \end{array}
\]
In terms of these variables and neglecting terms higher order than quadratic the Hamiltonian becomes 
\[
H = -\sum_{i\neq j}J(i-j) -2 \sum_{i\neq j}(J(i-j)(a^+_ia_j-a^+_ia_i) +O(a^4)
\]
Treating the above form of the Hamiltonian as defining a lattice Laplacian one solves for the eigenfunctions and eigenvalues through the equation:
 \[
\sum\limits_j {(\left\langle {i|H|\left. j \right\rangle } \right.}  - \delta _{i.j} \omega _\ell  )C_j^\ell   = 0
\]
The simplest case is the free or discrete-Neumann boundary condition where the spin on the boundary can rotate freely. In this case the eigenfunctions are given by
\[
\begin{array}{l}
C_i^\ell   = \mu_\ell \cos (\frac{{\pi \ell _1 }}{{2N_x }}(i_1  - \frac{1}{2}))\cos (\frac{{\pi \ell _2 }}{{2N_y }}(i_2  - \frac{1}{2}))\cos (\frac{{\pi \ell _3 }}{{2N_z }}(i_3  - \frac{1}{2}))\\
\end{array}
\]
where
\[
\begin{array}{l}
 \ell _1  = 0,1,2, \ldots ,2N_x  - 1 \\
 \ell _2  = 0,1,2, \ldots ,2N_y  - 1 \\
 \ell _3  = 0,1,2, \ldots ,2N_z  - 1 \\
 \end{array}
\]
and
\[
\mu_\ell = (\frac{2}{{2N_x (1 + \delta _{\ell _1 ,0} )}})^{1/2} (\frac{2}{{2N_y (1 + \delta _{\ell _2 ,0} )}})^{1/2} (\frac{2}{{2N_z (1 + \delta _{\ell _3 ,0} )}})^{1/2}
\]
The eigenvalues are given by:
\begin{equation}
\omega_\ell  = 2J\left(8 - 8\cos \frac{{\pi \ell_1 }}{{2N_x}}\cos \frac{{\pi \ell_2 }}{{2N_y}}\cos \frac{{\pi \ell_3 }}{{2N_z}}\right)
\end{equation}
The eigenfuction analysis is similar to \cite{Ryu} except that in that reference discrete Dirichlet boundary conditions and eigenfunctions given by sines were used.

Given these basis functions one can transform the $a_i$ variables to a new variable $b_\ell$ via:
\[
a_i  = \sum\limits_k {C_i^\ell b_\ell }.
\]
In terms of the tranformed variables the Hamiltonian becomes:
\[
H = \sum\limits_\ell  {\omega _\ell  } b_\ell ^ +  b_\ell   + O(b^4 ) +  \ldots ,
\]
and the magnetization is given by:
\begin{equation}
\frac{{M(T)}}{{M(0)}} = 1-\frac{1}{{NS}}\sum\limits_\ell {\langle b_\ell^+b_\ell\rangle }  = 1-\frac{1}{{NS}}\sum\limits_\ell {\frac{1}{{\exp (\omega _\ell /T) - 1}}} .
\end{equation}

The Bloch $T^{\frac{3}{2}}$ law for bulk iron follows from treating $\ell$ values  as continuous with magnitude from zero to infinity. 
For sufficiently small magnets, the compact and discrete aspects of the  magnon dispersion relation and finite size modifications to the Bloch law can be observed. 

Given the form of the magnetization it is useful to define an intrinic temperature for the bcc iron system given by $T_a  = D/a^2 = 2J/k_b = 395.902K$ where $D$ is the spin stiffness coeficient given by $D = .281eV{\AA}^2$ \cite{Crespo}. The lattice spacing for bcc iron is  $a = 2.87\AA = .287 \rm{nm}$.

In Fig. 2 we plot the quantity $M(T)/M(0)$ using (2.1) for different size magnets. The basic effect is that the magnetization increases for a given temperature if we decrease the size of the nanomagnet. The behavior is similar to the magnetization of nanoscale iron islands studied in \cite{Senz}. The finite size effect is larger as one goes to higher temperatures. Beyond 400K the free magnon approximation breaks down and one has to include magnon interactions.

For nanomagnets of size greater than $(28.7 \rm{nm})^3$ or 100 atoms on a side the finite size effects are small and one can treat them using a $1/N_x$ expansion. For the case $N_x=N_y=N_z$ the parameter $1/N_x$ completely determines the size of the nanomagnet. It is clear from Fig. 3 that there is a decrease in magnetization per atom as the size increases. For large values of $1/N_x$  the curve can be empirically fit to the formula:

\begin{figure}[htbp]
 
   \centerline{\hbox{
   \epsfxsize=5.0in
   \epsffile{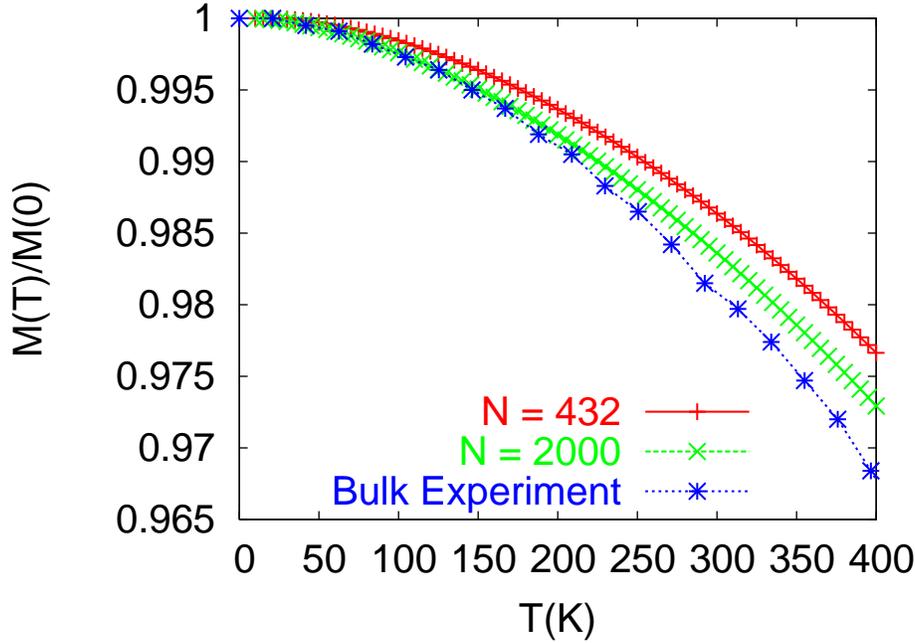}
     }
  }
  \caption{Finite size effects in iron nanomagnets. Calculated $M(T)/M(0)$ for bcc iron nanomagnets using (2.1). The nanomagnets are of size $ (1.722 \rm{nm})^3$ with 432 atoms and $(2.87 \rm{nm})^3$ with 2000 atoms. The experimental magnetization curve for bulk iron [3] is shown for reference.}
           
  \label{fig2}
 
\end{figure}

\begin{equation}
\frac{M(T)}{M(0)} =1 - .031245(T/T_a)^{3/2} + .056431(T/T_a)\frac{1}{N_x} + O(\frac{1}{N_x^2}) 
\end{equation}
Note as $M(0)$ is proportional to volume, the first and second terms in the above fit can be interpreted as volume and surface area contributions to the magnetization. The curve is nearly  linear versus $1/N_x$ in the range between $N_x = 600$ and $N_x =100$. We also studied the behavior between $N_x = 100$ and $N_x = 10$ (see Fig. 4) and also found that the  same  linear dependence is valid. This disagrees with the analysis of \cite{Crespo} [4th formula in Appendix B] who found that the leading finite size correction was proportional to $(T/T_a)^{5/4} \frac{1}{N_x^{1/2}}$. A possible source of error in the analysis of this correction is due to the use of the integral transformation
\[
\int_{\frac{2\pi}{N_x}}dkk^2 = \frac{1}{2(D/k_bT)^{3/2}} \int_{\frac{2\pi}{N_x}\frac{D}{k_bT}}dx x^{1/2}
\]
instead of the transformation 
\[
\int_{\frac{2\pi}{N_x}}dkk^2 = \frac{1}{2(D/k_bT)^{3/2}} \int_{(\frac{2\pi}{N_x})^2 \frac{D}{k_bT}}dx x^{1/2}
\]
with $x = \frac{Dk^2}{k_bT}$.

\begin{figure}[htbp]
 
   \centerline{\hbox{
   \epsfxsize=5.0in
   \epsffile{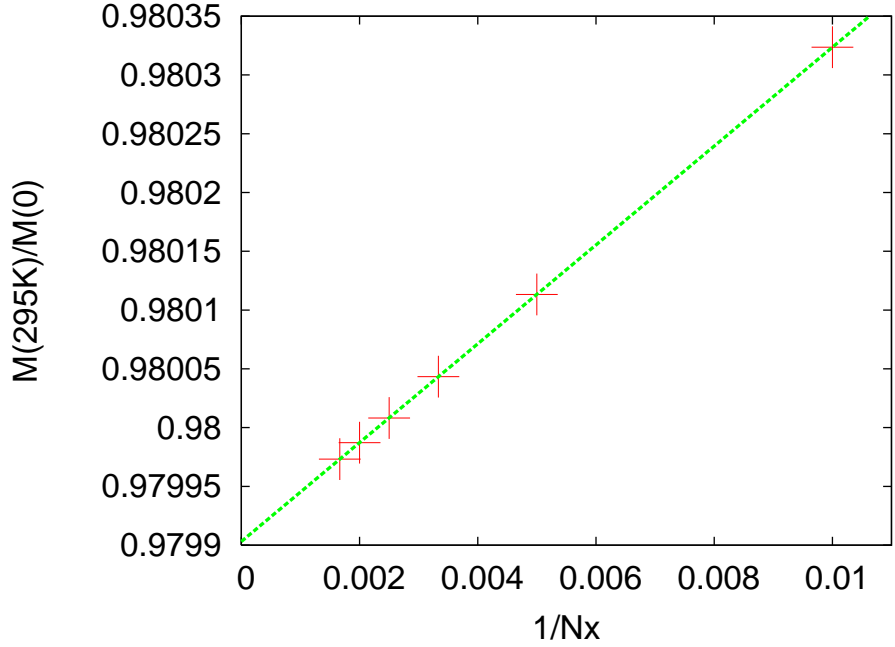}
     }
  }
 \caption{Calculated $ M(T=295K)/M(0)$ for bcc iron nanomagnets as a function of $1/N_x$, the reciprocal of the number of atoms on a side under the condition $N_x=N_y=N_z$ for the range $N_x = 600$ to $N_x = 100$.}

  \label{fig3}
 
\end{figure}

\begin{figure}[htbp]
 
   \centerline{\hbox{
   \epsfxsize=5.0in
   \epsffile{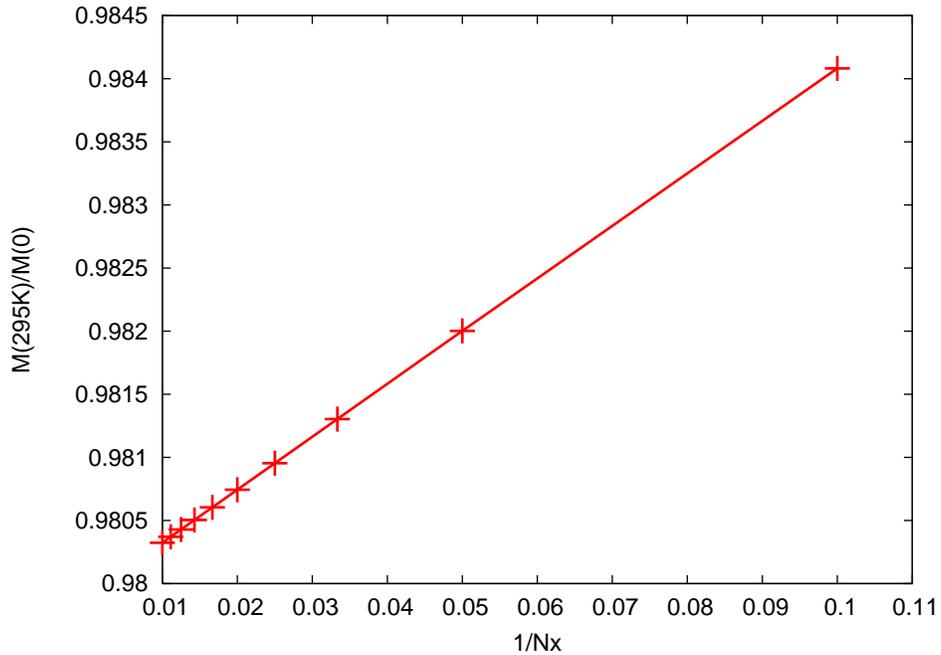}
     }
  }
 \caption{Calculated $ M(T=295K)/M(0)$ for bcc iron nanomagnets as a function of $1/N_x$, the reciprocal of the number of atoms on a side under the condition $N_x=N_y=N_z$ for the range $N_x = 100$ to $N_x = 10$.}

  \label{fig4}
 
\end{figure}

For $N$ large one should approach the magnetization curve of bulk iron. In Fig. 5 we plot the magnetization calculated using the magnon approximation for an iron nanomagnet with $639\times 639 \times 639$ atoms on a side, with volume $(183nm)^3$ and $521,834,238$ atoms. It is clear from the figure that additional effects such as spin wave interactions are physically important beyond temperatures of $400K$.

\begin{figure}[htbp]
 
   \centerline{\hbox{
   \epsfxsize=5.0in
   \epsffile{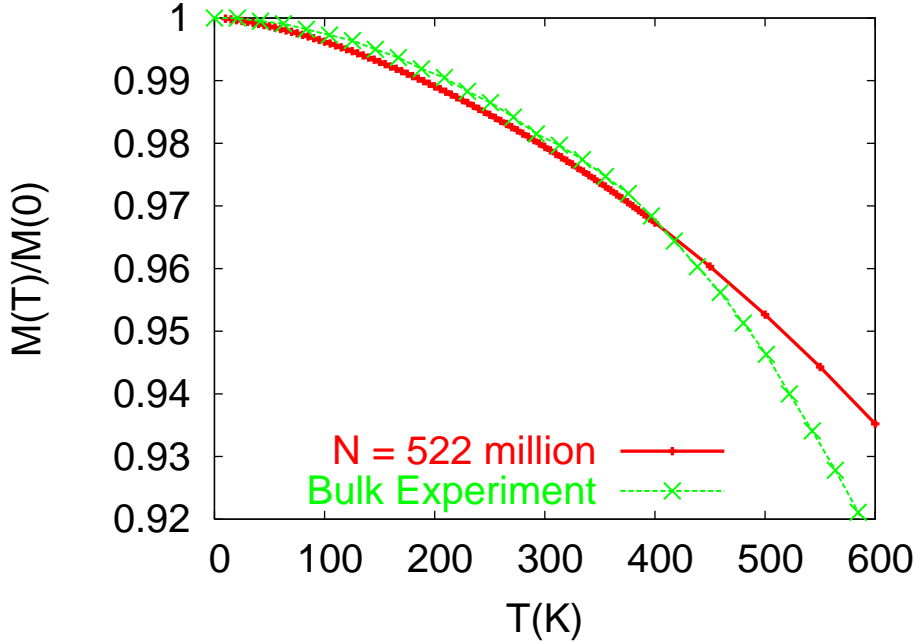}
     }
  }
 \caption{Magnon calculation of the magnetization for bcc iron nanomagnets with $639\times 639 \times 639$ atoms on a side, with volume $(183nm)^3$ and approximately 522 million atoms compared to bulk experimental curve of [3].}

  \label{fig5}
 
\end{figure}

\section{Shape effects in bcc iron nanomagnets}

One of the defining characteristics of a nanomaterial is that nanomaterial properties differ from those of the bulk. In the previous section we studied the finite size effect on the magnetization at fixed shape. In this section we fix the volume and study the effect of different shapes on the magnetization. Although we only consider shapes of the form of a parallelepiped with $N_x \times N_y \times N_z$ atoms on a side, it is possible to consider more complicated shapes by modifying  the discrete Laplacian which leads to the magnon dispersion relation.   

For the simple case when two of the lengths of the parallelepiped are equal say $N_x = N_y$ there is only one independent shape parameter which we take to be $h = N_z/N_x$ using the notation of \cite{Crespo}. At fixed volume the magnetization only depends on this parameter and temperature. In Figure 6 we plot the magnetization as a function of temperature for various shapes under the assumption of $N_x = N_y$. The dramatic dropoff in magnetization of the nanowire is in accord with the expectation of the absence of ferromagnetism in one and two dimensions \cite{Mermin2} regulated by the finite size of the magnet. An infinite one dimensional magnet would have zero magnetization at any finite temperature. Indeed the one dimensional quantum Heisenberg model was exactly solved in \cite{Bethe} and exhibits an absence of ferromagnetism.

To further illustrate this phenomenon we plot the magnetization in Fig. 7  at $100K$ and $295K$ as a function of the shape parameter $h$ with total volume $(34.44 nm)^3$. The data can be fit to the equation:
\begin{equation}
\begin{array}{l}
\frac{M(T)}{M(0)} =1 - .031245(T/T_a)^{3/2} + .056431(T/T_a)\frac{1}{N_m}\\
 + .03831(T/T_a)\frac{1}{N_m}(2h^{1/3}+h^{-2/3} -3) - .02083(T/T_a)\frac{1}{N_m}(2h^{-2/3}+h^{4/3} -3) 
\end{array}
\end{equation}
Here we have defined $N_m = (N_xN_yN_z)^{1/3}$. This equation reduces to (2.3) for $N_x=N_y=N_z$ and $h=1$. 

To gain further understanding of the various terms in this equation consider the more general case of a nanoribbon where two lengths need not be equal and two aspect ratios are defined by $h_1=N_z/N_x$ and $h_2=N_z/N_y$. In Figure 8 we plot nearly 2000 different shapes and fit to the equation:
\begin{equation}
\begin{array}{l}
\frac{M(T)}{M(0)} =1 - .031245(T/T_a)^{3/2} + .056431(T/T_a)\frac{1}{N_m}
\\
 + .03831(T/T_a)\frac{1}{N_m}(h_1^{2/3}h_2^{-1/3} +h_2^{2/3}h_1^{-1/3}+h_1^{-1/3}h_2^{-1/3}-3)\\
 - .02083(T/T_a)\frac{1}{N_m} (h_1^{-4/3}h_2^{2/3} +h_2^{-4/3}h_1^{2/3}+h_1^{2/3}h_2^{2/3}-3)
\end{array}
\end{equation}
Although complicated, this formula can be understood by considering an expansion of the magnetization before division by $M(0)$. Defining $L_x = aN_x$, $L_y = a N_y$, $L_z = a N_z$ and $\beta = 1/kT$, the second term on the right side is of the form $L_x L_y L_z/(D\beta)^{3/2}$ and is responsible for the usual $T^{3/2}$ law. The remaining terms are linear combinations of three types: $O_1 = (L_x L_y L_z)^{2/3}/(D\beta)$, $O_2 = (L_x L_y +L_y L_z + L_z L_x)/(D\beta)$ and $O_3 = (L_x^2 +L_y^2 + L_z^2)/(D\beta)$. These are the unique terms which are quadratic in the $L$'s and invariant under permutations of the axes of the parallelepiped. Expressing the $L$'s in terms of the parameters $N_m$, $h_1$ and $h_2$ and dividing by $M(0)$ (which is proportional to $N_x N_y N_z$) we obtain an expression of the form (3.2). Equation (3.2) reduces to (3.1) when $h_1 = h_2$.  As before (3.1) reduces to (2.1) when $h=1$.

\begin{figure}[htbp]
 
   \centerline{\hbox{
   \epsfxsize=5.0in
   \epsffile{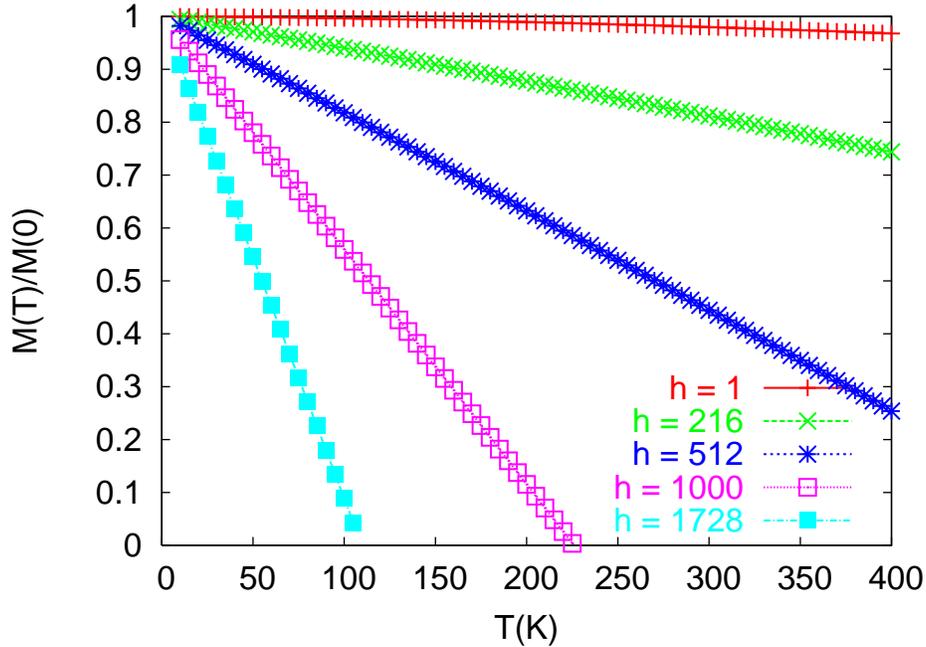}
     }
  }
 \caption{Shape effects in nanomagnets. $ M(T)/M(0)$ for bcc iron nanomagnets of different shape, each consisting of 3,456,000 atoms with total volume  $(34.44 \rm{nm})^3$. The shape from top to bottom are $120 \times 120 \times 120$; $20 \times 20 \times 4320$; $15 \times 15 \times 7,680$; $12 \times 12 \times 12,000$ and $10 \times 10 \times 17,280$ atoms on a side.}

  \label{fig6}
 
\end{figure}

\begin{figure}[htbp]
 
   \centerline{\hbox{
   \epsfxsize=5.0in
   \epsffile{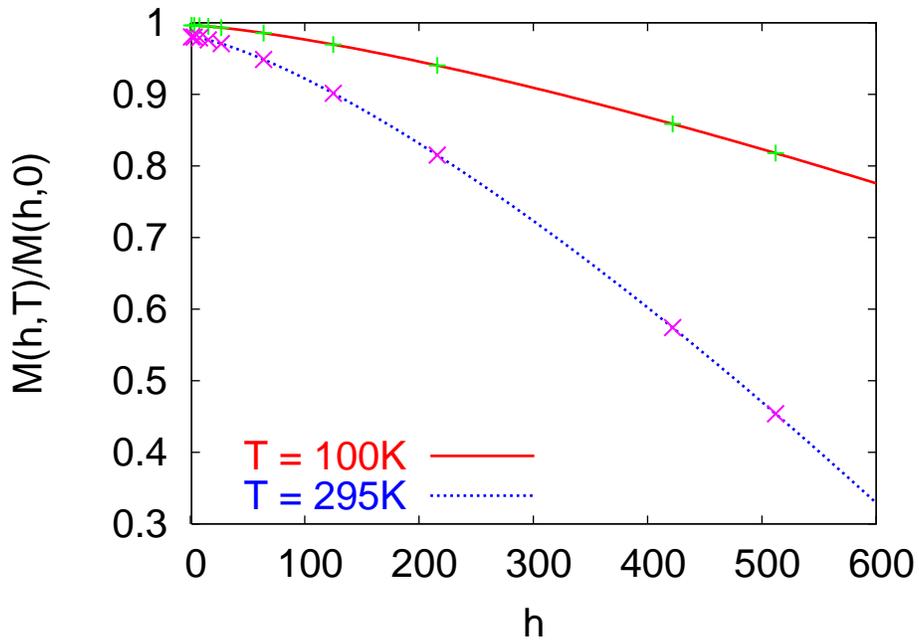}
     }
  }
 \caption{Dependence of magnetiization on the shape parameter of a magnetic nanowire. We calculated $ M(T=100K)/M(0)$ and $M(T=295K)/M(0)$ for bcc iron nanomagnets of different aspect ratios $h = N_z/N_x$ with $N_x = N_y$  each consisting of 3,456,000 atoms with fixed total volume  $(34.44 \rm{nm})^3$. The smooth curves are a numerical fit to the form (3.1).}

  \label{fig7}
 
\end{figure}

\begin{figure}[htbp]
 
   \centerline{\hbox{
   \epsfxsize=5.0in
   \epsffile{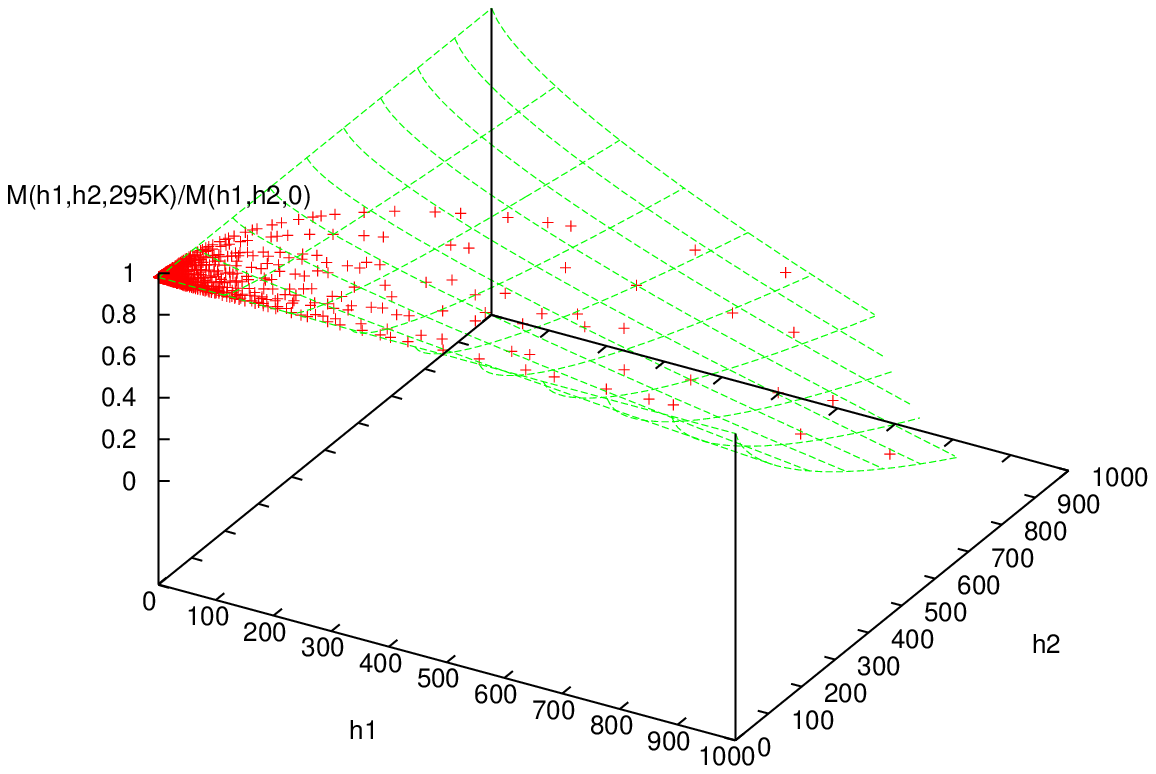}
     }
  }
 \caption{Dependence of magnetization on the shape parameters of a magnetic nanoribbon. We calculated $ M(T=295K)/M(0)$ for bcc iron nanomagnets of nearly 2000 different aspect ratios $h_1 = N_z/N_x$ and $h_2 = N_z/N_y$  each consisting of 3,456,000 atoms with fixed total volume  $(34.44 \rm{nm})^3$. The smooth surface is a numerical fit of the form (3.2).}

  \label{fig8}
 
\end{figure}

An experimental determination of the magnetization of iron nanoribbons was carried out in \cite{Crespo}. In that work the long direction of the nanoribbon was not specified. As in that work we use the spin-wave model to compare with this data and fit the long direction of the nanomagnet. The prediction of the spinwave model for the nanoriboon is shown in Fig. 9. Our determination of the length of the long direction is 3645 nm. This is of the same order of magnitude as the  2000 nm determination in \cite{Crespo} where the spin wave method, but not the bcc lattice finite size dispersion relation was used.

\begin{figure}[htbp]
 
   \centerline{\hbox{
   \epsfxsize=5.0in
   \epsffile{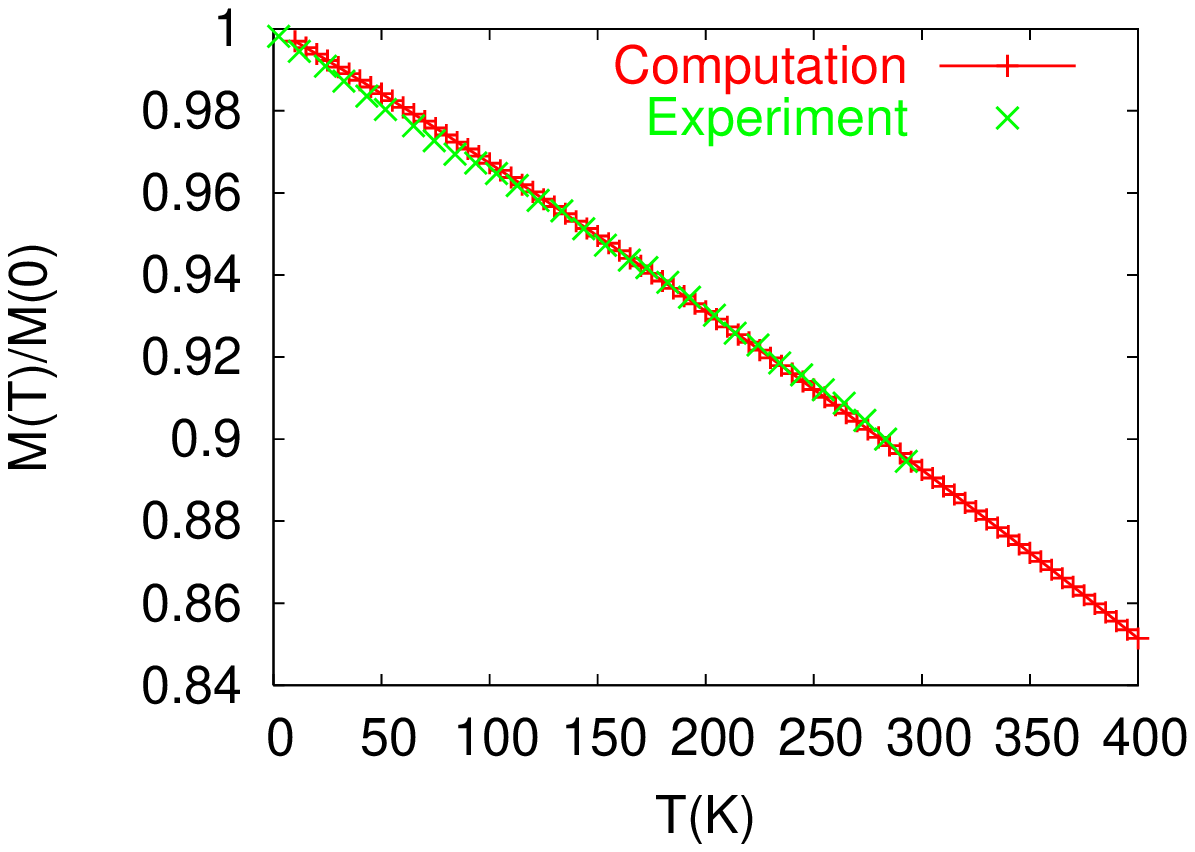}
     }
  }
 \caption{Magnetization curve for an iron nanoribbon with $14 \times 167 \times 12,700$ atoms on a side. The total number of atoms was 59,385,200. The length of the long direction was determined from a fit to the data of [7] to be  3645 nm.}
     
  \label{fig9}
 
\end{figure}

\section{Conclusion and Future Direction}

We developed and validated an efficient computational approach to nanomagnets. The method takes input from a first principle calculation of the Heisenberg model parameters and uses a magnon approximation to the quantum Heisenberg model to calculate the magnetic properties. Although we demonstrated the method for iron, this type of calculation can be carried out for other materials and remains valid at room temperature. Composite systems such as iron-rhodium can be treated by our method as first principle calculations of the $J$ values are known \cite{Ant3} in that case.

\section*{Acknowledgements}
This manuscript has been authored in part by Brookhaven Science Associates, LLC, under Contract No. DE-AC02-98CH10886 with the U.S. Department of Energy.

\end{document}